\documentclass[aps, pra, showpacs, twocolumn, amsfonts, amsmath, amssymb, superscriptaddress, floatfix]{revtex4}

\usepackage{amsmath}
\usepackage{graphicx}
\begin{document}
\newcommand{\be}{\begin{equation}}
\newcommand{\en}{\end{equation}}
\newcommand{\bal}{\begin{align}}
\newcommand{\eal}{\end{align}}
\newcommand{\dw}{\Delta \omega}
\newcommand{\om}{\omega}
\newcommand{\pa}{\partial}
\newcommand{\hb}{\hbar}
\newcommand{\ef}{\'{e} }
\newcommand{\af}{\`{a} }

 \title{Localization of an inhomogeneous Bose-Einstein condensate in a
   moving random potential}

\author{Ardavan Alamir}

\affiliation{Universit\'e de Nice - Sophia Antipolis, Institut non Lin\'eaire de Nice, CNRS, 1361 route des Luciole
s, 06560 Valbonne, France}

\author{Pablo Capuzzi} 

\affiliation{Departamento de F\'{\i}sica, Facultad de Ciencias Exactas
  y Naturales, Universidad de Buenos Aires, 1428, Buenos Aires,
  Argentina} 

\affiliation{Instituto de F\'{\i}sica de Buenos Aires --
  CONICET, Argentina} 

\author{Patrizia Vignolo}

\affiliation{Universit\'e de Nice - Sophia Antipolis, Institut non
  Lin\'eaire de Nice, CNRS, 1361 route des Luciole s, 06560 Valbonne,
  France}

\begin{abstract}
  We study the dynamics of a harmonically trapped
  quasi-one-dimensional Bose-Einstein condensate subjected to a moving
  disorder potential of finite extent. We show that, due to the
  inhomogeneity of the sample, only a percentage of the atoms is
  localized at supersonic velocities of a random potential. We find
  that this percentage can be sensitively increased by introducing
  suitable correlations in the disorder potential such as those
  provided by random dimers.

\end{abstract}

 
\maketitle
\section{Introduction}
Suppression of wave transport in non-dissipative linear systems can be
induced by the presence of disorder: the scattered waves from the
modulation of the random disorder potential destructively interfere in
the forward direction, with a resulting vanishing wave transmission.
This phenomenon is called Anderson localization (AL)
\cite{Ande58,Ande85}.  In three dimensions (3D), AL takes place for
states with energy less than a threshold (mobility edge). In two (2D)
and one dimensions (1D) and in the absence of interactions, all
single-particle quantum states are expected to be localized
\cite{Ande79}. However, in the presence of correlations, the situation
differs and a subset of delocalized states can appear in the spectrum
\cite{Phil90} or an effective
\cite{Tessieri2002,Kuhl08,Gurevich2009,Luga09,Piraud2011} or even a
true \cite{deMour98} mobility edge can be observed even at low
dimensions.

Anderson localization of noninteracting atomic matter waves was
observed in momentum and real space. In momentum space using
kicked-rotor setups in 1D \cite{Moore1995} and 3D \cite{Chabe2008},
while in real space in 1D~\cite{Billy2008,Roati2008}, and very
recently in 3D~\cite{Kondov2011,Jendrzejewski2012}.  On the other
hand, in 2D only anomalous diffusion has been observed
\cite{Saintvincent2010}. In the last few years it has been
experimentally demonstrated that disorder strongly affects the
transport properties and dynamics of a BEC, as for instance
illustrated in \cite{Hulet2008}.

One of the outstanding challenges of physics is to understand the
interplay between disorder and interactions.  In the case of an
interacting condensate, wave scattering from the random potential does
not occur if the wave group velocity is lower than a critical velocity
$v_c$ that coincides with the (local) sound velocity $c$ in the limit
of small random potential amplitude and decreases down to vanishing
values in the strong disorder limit \cite{Ianeselli2006}.  Thus, to
observe AL in an interacting BEC it is necessary that the speed of the
relative motion between the superfluid and the disorder potential is
larger than $v_c$. This setup was proposed and theoretically studied
in Refs.  \cite{leboeuf07,leboeuf09,Albert2010}. These authors studied
the flow of an homogeneous quasi-one-dimensional Bose-Einstein
condensate through a disorder potential of finite extent. That
disorder potential moves with a velocity $v$ with respect to the
condensate. In the subsonic regime, the flow is superfluid and the
density profile is stationary. In the opposite supersonic regime, a
region of stationary flow also exists, but in this case energy
dissipation occurs. In this domain, depending on the extent of the
disorder potential, the system is either in a Ohmic or in an AL
regime, respectively characterized by a transmission decreasing
linearly or exponentially with the size of the system $L$.

At variance of Refs. \cite{leboeuf07,leboeuf09,Albert2010}, in this
paper we study the effects of the inhomogeneity of a cigar-shaped
trapped BEC in the presence of a moving disorder potential.  We
investigate the possibility of observing BEC localization by looking
at the position of the center of mass of the condensate. If the center
of mass moves along with the moving potential then the system shall be
in the AL regime or in another kind of localized phase. Because of the
inhomogeneity, we observe the localization of only a percentage of the
atoms in the BEC.  This percentage of localized atoms can be increased
or suppressed by introducing ad-hoc short-range correlations on the random
potential.

The paper is organized as following. In Sec. \ref{themodel} we
introduce the time-dependent nonpolynomial nonlinear Schr\"odinger
equation (NPSE) that describes the condensate dynamics in the
elongated geometry and in the presence of a moving disorder potential
that we characterize by its auto-correlation function.  As discussed
and shown in Sec. \ref{secloc}, the disorder potential drags the atoms
with an efficiency that depends on both the scattering properties of
each impurity and on the impurity density.  The case of two types of
impurities, single and dimerized, at different densities, have been
studied, highlighting the role of correlations in the localization
dynamics.  Because of the inhomogeneity of the BEC, we observe that
the drag force is more efficient in the BEC tails where the local sound
velocity is lower and superfluidity breaks down at small drift
velocities $v$.  Our concluding remarks are given in Sec. \ref{concl}.
\section{The model}
\label{themodel}
\subsection{Equation of motion for the BEC wavefunction}
\noindent Our starting point is the equation of motion for a 3D BEC
trapped into a cigar-shaped potential. Such an equation is known as
the Gross-Pitaevskii equation (GPE) \be
\label{3dgpe}
i \hbar \frac{\partial}{\partial t} \psi (\textbf{r},t) = \left[ -
  \frac{\hbar^2}{2 m} \nabla^2 + U(\textbf{r}) + g N
  |\psi(\textbf{r},t)|^2 \right] \psi(\textbf{r},t) .  \en The wave
function $\psi(\textbf{r},t)$ describes the condensate, which is
constituted of $N$ atoms of mass $m$. $g=4 \pi \hbar^2 a_s / m$ stands
for the interaction coupling constant with $a_s$ being the {\it
  s}-wave scattering length between atoms; in our system, interatomic
interactions are repulsive and, as a result, $a_s > 0$. The trapping
potential $U(\textbf{r})$ is given by the sum of a static cigar-shaped
harmonic trap and a time-dependent random potential: \be
\label{potential1}
U(\textbf{r},t) = \frac{1}{2} m \omega^2_\perp (x^2 + y^2) +
\frac{1}{2} m \omega^2_z z^2 + V(z,t) \en with $\omega_{\perp}$ and
$\omega_{z}$ the trapping frequencies in the perpendicular and
longitudinal directions, respectively and $\omega_z \ll
\omega_\perp$. The last time-dependent term in (\ref{potential1})
corresponds to a random potential that is fixed in the moving frame
$z'=z-vt$, ${\bf v}=v\hat e_z$ being the drift velocity.

Under this trap geometry and a further assumption discussed below, the
3D GPE can be reduced to an effective 1D time-dependent
NPSE~\cite{salasnich02}. The advantage of the 1D NPSE is that it is
easier to deal with when making computations. In order to obtain such
dynamical equation, we begin with a variational Ansatz \be
\psi(\textbf{r},t) = f(z,t) \phi(\textbf{r},t)=f(z,t)
\frac{e^{-(x^2+y^2) / 2 \sigma^2(z,t)} }{\sqrt{\pi} \sigma(z,t)} \en
where the transverse part $\phi(\textbf{r},t)$ is modeled by a
Gaussian function with variance $\sigma(z,t)$. The validity of this
description is based on the assumption that $\sigma(z,t)$ slowly
varies as a function of $z$ and $t$ such that the kinetic energy term
$\pa^2 / \pa z^2$ associated to $\phi(\textbf{r},t)$ can be neglected.
Both the longitudinal wave function $f(z,t)$ and the variance
$\sigma(z,t)$ are determined by the energy variational principle. For
$f(z,t)$ one gets the NPSE \be
\begin{split}
i \hbar \frac{\partial}{\partial t} f &= \left[ - \frac{\hbar^2}{2 m} \frac{\pa^2}{\pa z^2} + \frac{1}{2} m \om^2_z z^2 + V(z,t)\right.\\
&\left. + \hbar \om_{\perp} \frac{1 + 3 a_s N |f|^2 }{\sqrt{1 + 2 a_s N |f|^2}} \right] f.
\end{split}
\en
The variance is given by 
\be\label{13071}
\sigma^2 (z,t) = a^2_\perp\sqrt{1 + 2 a_s N |f(z,t)|^2},
\en
where $a_\perp = \sqrt{\hbar / (m \omega_\perp)}$ is the oscillator length in the transverse direction.
The 3D density profile and velocity field are then	 
\begin{align}
\rho (\textbf{r}) & = \tilde{\rho} (z) \frac{e^{-r^2 / \sigma^2}}{\pi \sigma^2}, \label{thur07071} \\
\textbf{v} (\textbf{r}) & = v (\textbf{r}) \hat{z} = \frac{\hb}{2 m i} \frac{f^{' *}(z) f(z) - f'(z) f^*(z)}{\tilde{\rho}} \hat{z},
\end{align}
with $\tilde{\rho} (z) = |f|^2$ the integrated 1D density. 

The NPSE is numerically solved using a split-step method and spatial
Fast Fourier transforms (FFT). First we compute the equilibrium
density profile in the presence of a static disorder potential. Then,
we switch on the drift velocity $v$ and compute the time evolution of
the condensate wavefunction $f(z,t)$. In this work, we focus on a
system of $10^5$ 87-Rubidium atoms subject to a transverse confinement
of $\omega_\perp = 2 \pi \times 500$ Hz and a longitudinal
confinement of $\omega_z = 2 \pi \times 7$ Hz. The $s$-wave
scattering length has been fixed at $a_s = 80$ Bohr radii.

\subsection{The random potential}
\label{subsecthepot}
\begin{figure}
\includegraphics[width=1\linewidth]{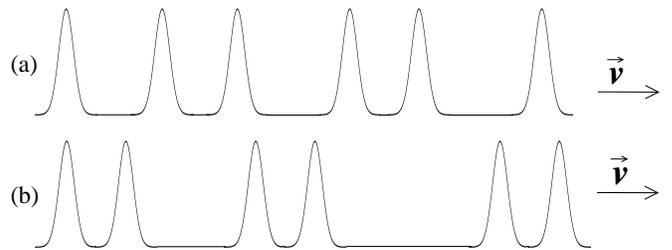}
\caption{\label{fig1}Schematic representation of the disorder
  potential. (a) The Gaussian peaks are randomly distributed (AM). (b)
  The peaks are distributed in a random dimer sequence (RDM).}
\end{figure}
The random potential $V(z,t)$ is modeled by the sum of
$N_{\textmd{dis}}$ Gaussian functions of height $V_{\textmd{dis}}$ and
width $w$, randomly distributed at positions $z_i=j_id$, where $j_i$
is a random integer number and $d$ fixes the minimal distance between
the peaks. Such a disorder potential could be realized by deeply
trapping some impurities (heavy atoms of another species) in an
optical lattice strongly detuned from the condensate atomic
frequencies \cite{Gavish2005a,Vign03,schaff10}.  When the disorder
pattern is pulled with a constant speed $v$ through the system, if $v$
is lower than the sound speed $c=\sqrt{\mu/(2m)}$~\cite{zaremba98}, we
expect the disorder not to affect the system because of the superfluid
nature of the gas itself \cite{leboeuf07}.  In contrast, the effect of
the disorder potential should appear at $v\gtrsim c$ where the kinetic
energy starts to compete with the interaction energy and the limit of
a noninteracting gas is reached for $v\gg c$.

\begin{figure}
\includegraphics[width=0.9\linewidth]{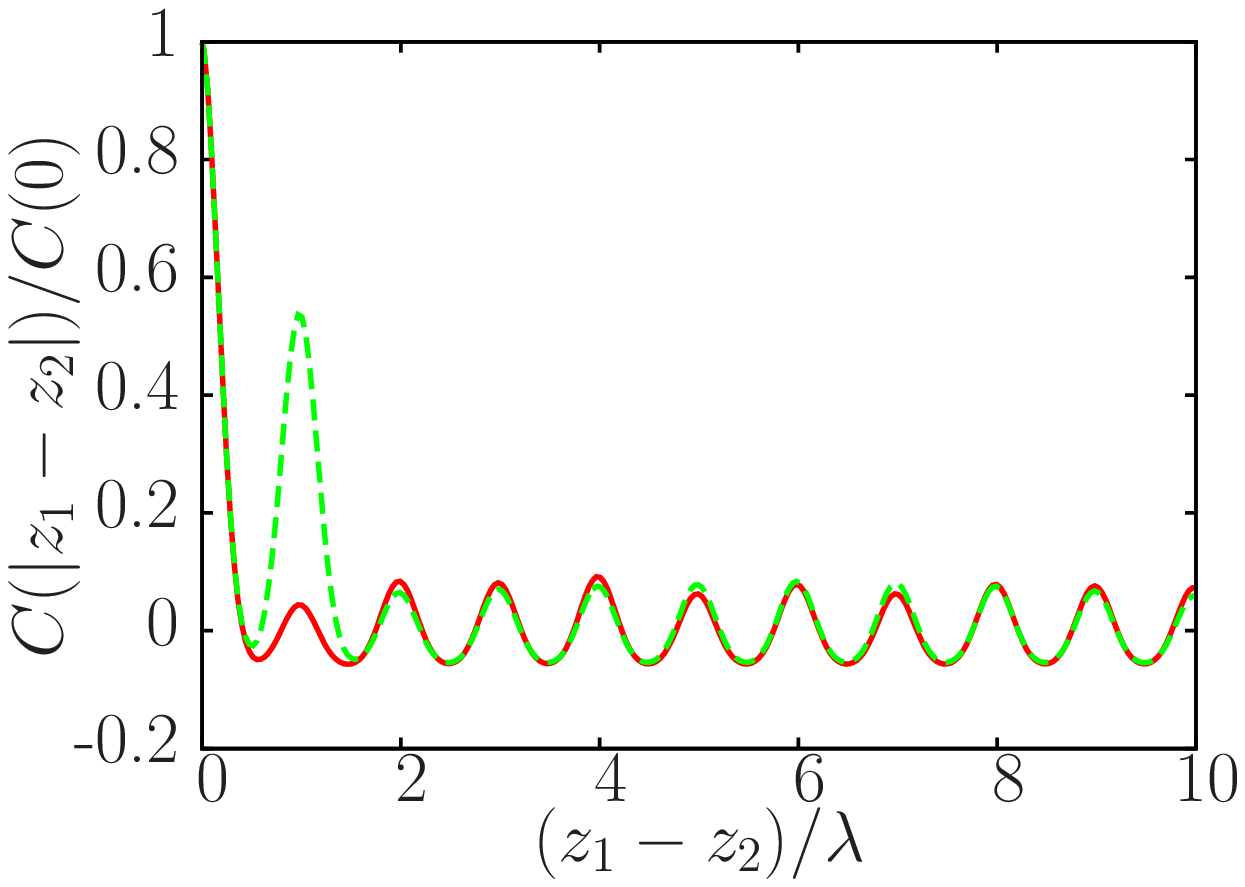}
\includegraphics[width=0.9\linewidth]{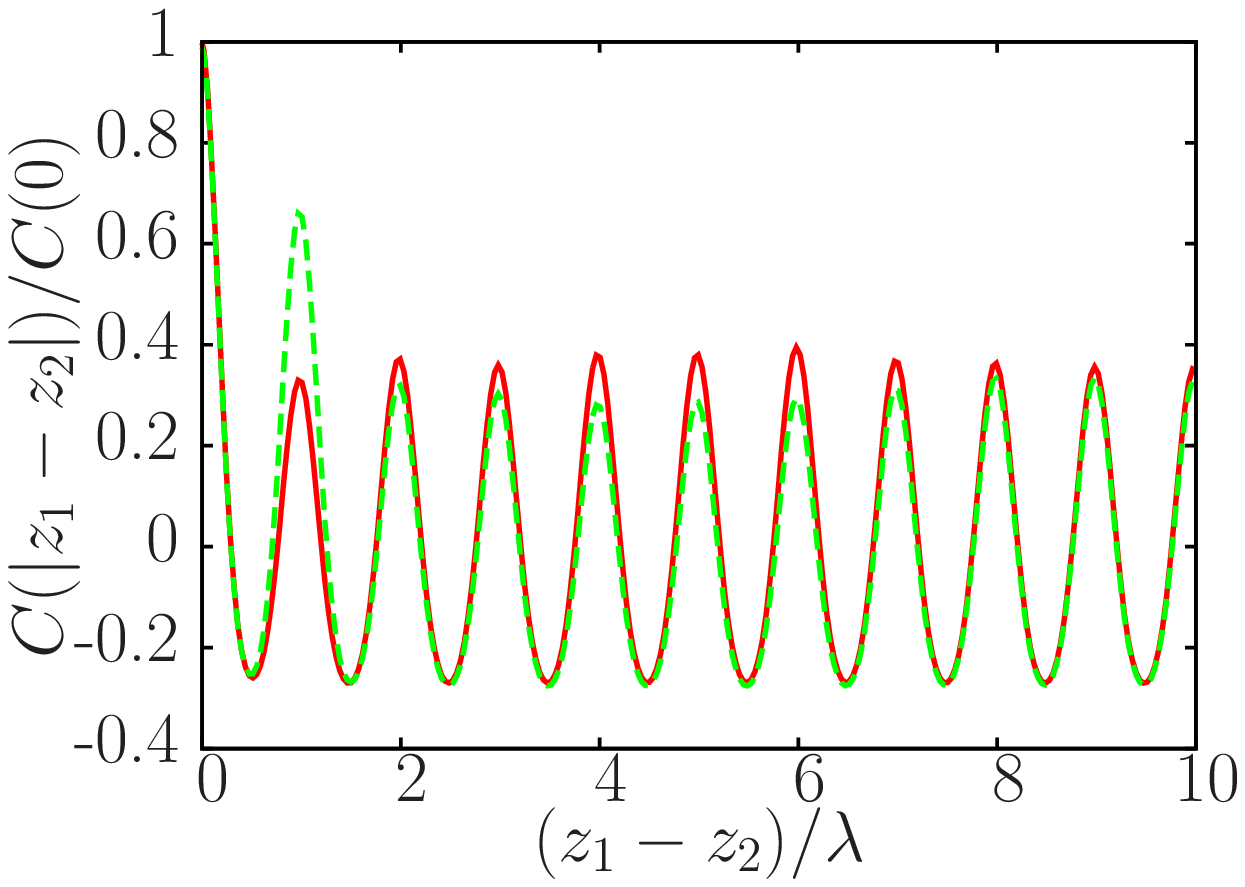}
\caption{\label{fig-autocorr}(Color online) Rescaled auto-correlation function 
$C(|{z_1}-{z_2}|)/C(0)$ as a function of $(z_1-z_2)/\lambda$
for the AM (red continuous line) and RDM 
(green dashed line) for the case $d=\lambda$. The top panel corresponds to
$n_{\textmd{\textmd{dis}}}=0.12\lambda^{-1}$, and the bottom panel to $n_{\textmd{\textmd{dis}}}=0.5\lambda^{-1}$.}
\end{figure}

We will consider two sorts of potential patterns: (i) an Anderson-like
distribution, that we will call Anderson Model (AM), where the $j_i$'s
are randomly distributed; (ii) a Random Dimer Model (RDM) distribution
where peaks are dimerized and the dimers are randomly distributed (see
Fig \ref{fig1}). The dimers are stamped by the peak-to-peak distance
$\ell$.  Hereafter we consider a disorder potential characterized by
an amplitude $V_{\textmd{dis}}=0.02 E_r$, where the recoil energy
$E_r=h^2/(2 m\lambda^2)$ for each atom of mass $m$ is defined with
respect to the wavelength $\lambda=780$ nm, characterizing the D2
hyperfine Rubidium transition.  The dimer peak-to-peak distance $\ell$
was set equal to $\lambda$ and the width of a single bump $w$ is fixed
at $140\textmd{nm}$, roughly $\lambda/5$, ensuring no sizeable overlap
between Gaussian functions.  The disorder potentials can be
characterized in term of their autocorrelation functions
\begin{equation}
C(|{z_1}-{z_2}|)=\langle (V({z_1},t)-\langle V({z_1},t)\rangle)
(V(z_2,t)-\langle V({z_2},t)\rangle)\rangle,
\end{equation}
which strongly influence the nature of the energy states for low
amplitude disorder near equilibrium.  In the case of the AM and
RDM potentials in Fig. \ref{fig-autocorr} we plot the autocorrelation
function for the case $d=\lambda$ and for two different peaks
densities: $n_{\textmd{\textmd{dis}}}=0.12\lambda^{-1}$ (top panel)
and $n_{\textmd{\textmd{dis}}}=0.5\lambda^{-1}$ (bottom panel). It is
worthwhile remarking than the peak density corresponds to the average
number of Gaussian peaks both in the AM and RDM potential and thus
in average the number of dimers in RDM is half the number of
peaks in the AM. 
The modulation with spatial period $\lambda$ for both the AM and the
RDM evinces that bumps and dimers are both randomly distributed over
discretized positions of step $d=\lambda$.  The main difference
between the AM and the RDM is that the RDM has a larger peak at
$z_1-z_2=\lambda$ because of the dimer structure with $\ell=\lambda$.
This is well visible at low peak density (top panel), while by
increasing the bump density the probability to find dimerized
structures in the AM potential increases as well. This is seen in the
heights of the peaks at $z_1-z_2=j\lambda$ for the two models becoming
closer even for $j=1$ (bottom panel).

\begin{figure}
\includegraphics[width=0.9\linewidth]{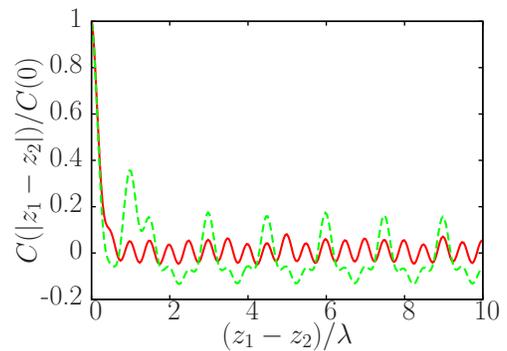}
\caption{\label{fig-autocorr2} (Color online) Same as in Fig.
  \ref{fig-autocorr} but for the case $d=\lambda/2$ and
  $n_{\textmd{\textmd{dis}}}=0.5\lambda^{-1}$.}
\end{figure}

In Fig. \ref{fig-autocorr2} we show $C(|z_1-z_2|)$ for the case 
$d=\lambda/2$ and $n_{\textmd{\textmd{dis}}}=0.5\lambda^{-1}$ (keeping fixed $\ell=\lambda$).
In this case the RDM peak at $z_1-z_2=\lambda$ is lower of more than a factor 2
with respect to the case with $d=\lambda$ at the same disorder density 
(bottom panel of Fig. \ref{fig-autocorr}): we thus expect 
the dimer structure to play a minor role for the case $d=\lambda/2$.

\section{The localization fraction and the drag force}
\label{secloc}
If the condensate or a part of it is localized, we expect it to follow
the pulled disorder potential.  The localized portion of the
condensate glues to the disorder "bandwagon" and therefore travels the
same distance as the disorder potential.  This dynamics depends on the
forces experienced by the atoms. The force acting on the BEC
center-of-mass has two terms, $F=F_{h}+F_{\textmd{dis}}$, one due to
the harmonic confinement, $F_{h}=-m\omega_z^2 z_{\mathrm{c.m.}}$ and
the other due to the disorder potential
\begin{equation}\label{020520122}
F_{\textmd{dis}}=-\int_{-\infty}^{+\infty} dz |f(z,t)|^2\dfrac{\partial V}{\partial z}.
\end{equation}
For small center-of-mass displacements $\Delta
z_{\mathrm{c.m.}}\simeq0$, the leading term is the drag force
$F_{\textmd{dis}}$ due to the disorder potential. In this regime the
localization fraction $N_{loc}/N$ can be deduced by the ratio between
the $\Delta z_{\mathrm{c.m.}}$ and the distance $\Delta
z_{\textmd{dis}}$ traveled by the disorder potential in the same time
interval, namely in this regime we can identify
\begin{equation}
\dfrac{N_{loc}}{N}=\dfrac{\Delta z_{\mathrm{c.m.}}}{\Delta z_{\textmd{dis}}}.
\label{nloc}
\end{equation}
Indeed
if the center of mass travels the same distance as the disorder
potential, this would mean that the 100\% of the atoms are localized.
The localized condensate will stop following the moving potential and it will 
change direction at a time $t_f$ at the position $z_f$ verifying
\begin{equation}\label{020520121}
\int_0^{t_f}F_{\textmd{dis}}(t)v_{\mathrm{c.m.}}(t) {\rm d}t=\dfrac{1}{2}m\omega_z^2z_f^2.
\end{equation} 
Thus, the turning point $z_f$ will provide a direct measure of the
average value of the drag force during the condensate forward motion
via the relation
\begin{equation}
z_f=2 \bar F_{\textmd{dis}}/(m\omega_z^2),
\label{eq:turningpoint}
\end{equation}
 where we have defined 
\begin{equation}
\bar F_{\textmd{dis}}=\int_0^{t_f}F_{\textmd{dis}}(t)v_{\mathrm{c.m.}}(t) {\rm d}t /z_f=
\int_0^{z_f}F_{\textmd{dis}}(t){\rm d}z_{\mathrm{c.m.}}/z_f.
\label{eq:av-dragforce}
\end{equation}
The dependence of the drag force with the drift velocity gives us a
direct measure of the loss of superfluidity in the system. According
to the Landau criterion \cite{Landau1941} a single impurity is
expected to flow without friction below a certain velocity,
corresponding to the sound velocity for a weakly interacting BEC.

\subsection{The single impurities}
\label{sec-corr}
\begin{figure}
\includegraphics[width=0.9\linewidth]{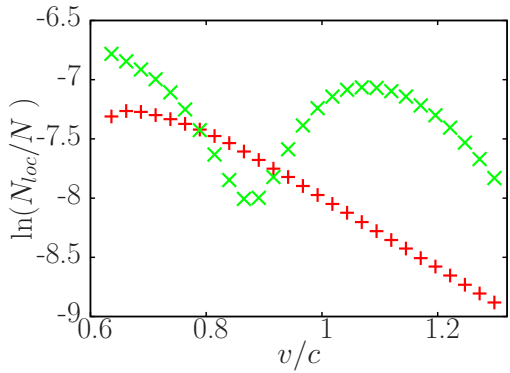}
\caption{\label{fig-singledef} (Color online) Localized BEC fraction
  (in logarithmic scale) as a function of the disorder potential drift
  velocity $v$, in units of the sound speed $c$ evaluated at the
  center of the trap. The red crosses correspond to a single bump and
  the green stars to a dimer.}
\end{figure}

The localization efficiency of a disorder pattern depends on the
impurity density and on the reflectivity of each impurity.  In this
work we are comparing single impurities randomly distributed with
dimerized structures.  With the aim to understand the difference in
behavior of the localization efficiency of a dimer with respect to a
single bump, we first look at the condensate dynamics in the presence
of a moving single bump (red crosses in Fig. \ref{fig-singledef}) and
of a moving dimer (green stars in Fig. \ref{fig-singledef}).

\begin{figure}
\includegraphics[width=0.5\linewidth]{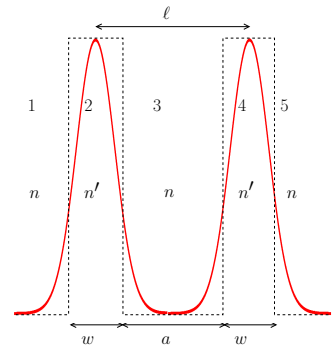}
\caption{\label{fig2}(Color online) Analogous optical system of the
  dimer structure present in our model.}
\end{figure}

\begin{figure}
\includegraphics[width=0.8\linewidth]{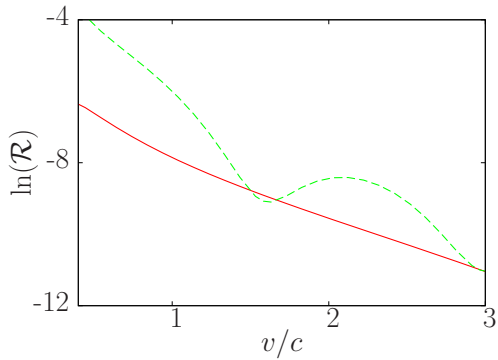}
\caption{(Color online) Reflectivity (in logarithmic scale) of a
  single defect (red continuous line) and of a dimerized structure
  (green dashed line) as a function of the disorder potential
  velocity\label{figrefl}.}
\end{figure}

In Fig. \ref{fig-singledef} we plot the fraction of atoms that follows the
moving defect over a distance of $133 \mu$m.
We observe that the localization fraction of the dimer is a strongly
non-monotonic function of $v$.  The single dimer localizes the atoms
less efficiently than a single bump for $v\simeq 0.9c$, while it is
more efficient by a factor of 3 over a velocity range of 1--1.2
$v/c$. The suppression and the enhancement of the localization are
both a signature of some interference effect due to the internal
structure of the single dimer.  This behavior can be qualitatively
reproduced by considering the analogous optical system of our model,
schematically shown in Fig. \ref{fig2}.  It consists of two dielectric
slabs of refraction index $n'$, width $w$, at distance $a=\ell-w$,
merged in a medium of refractive index $n$, with $n=1$, and $n'\simeq
1+V_{\textmd{dis}}/mv^2$ in the limit $mv^2\gg V_{\textmd{dis}}$ (and
$v>c$).
This model corresponds to associate to an incident wave of energy
$E=mv^2/2$ and wavevector $\kappa=mv/\hbar$ a transmitted wave of
wavevector $\kappa'=\kappa\sqrt{1-V_{\textmd{dis}}/E}\simeq
\kappa(1-V_{\textmd{dis}}/2E)$ in the regions where the disorder
potential is present.  The reflection coefficient for an incident wave
of wavevector $\kappa$ through the two-slab system (from region 1 to
region 5, as shown in Fig.  \ref{fig2}) can be written as
\begin{equation}
r_{15}=\dfrac{r_{12}+r_{25}e^{2i\alpha}}{1+r_{12}r_{25}e^{2i\alpha}}
\label{refl_tot}
\end{equation} 
with
\begin{equation}
\begin{split}
r_{25}=&\dfrac{r_{23}+r_{35}e^{2i\beta}}{1+r_{23}r_{35}e^{2i\beta}}\\
r_{35}=&\dfrac{r_{34}+r_{45}e^{2i\alpha}}{1+r_{34}r_{45}e^{2i\alpha}}\\
\end{split}
\end{equation}
where $\alpha=\frac{n}{n'}\kappa w$, $\beta=\kappa a$, $r_{34}=r_{12}=-r_{23}=-r_{45}=(n-n')/(n+n')$.
Equation (\ref{refl_tot}) takes the form
\begin{equation}
\begin{split}
&r_{15}=r_{12}\times\\
&\dfrac{\left[-1-2e^{2i\alpha}+e^{2i\beta}+r_{12}^2e^{4i\alpha}+r_{12}^2e^{2i(\alpha+\beta)}-2e^{2i(\beta+2\alpha)}\right]}
{1-2r_{12}^2e^{2i\alpha}+r_{12}^2e^{4i\alpha}+r_{12}^2e^{2i\beta}-2r_{12}^2e^{2i(\alpha+\beta)}+2r_{12}^4e^{2i(2\alpha+\beta)}}\\
\end{split}
\end{equation}
The behavior of the reflectivity $\mathcal{R}=|r_{15}|^2$ of the dimer
structure must be compared with that of a single bump
\begin{equation}
|r_{13}|^2=\left|\dfrac{r_{12}(1-e^{2i\alpha})}{1-r^2_{12}e^{2i\alpha}}\right|^2.
\end{equation}
This is shown in Fig. \ref{figrefl}. For our choice of the parameters,
the reflectivity of the dimer oscillates with respect to that of a
single bump that decreases monotonically, in qualitative agreement
with what observed for the localization fraction shown in
Fig. \ref{fig-singledef}. The shift of the minimum position for the
dimer reflectivity with respect to the localization may be attributed
to on the one hand, the Gaussian shape of bumps as
compared to the rectangular shape of the dielectric slabs, and on the
other hand to the inhomogeneity and non-linearity of the system. All
these factors are not taken into account in the current optical model.
Finally, let us remark that both the single bump and the single dimer
yield full delocalization ($\mathcal{R}=0$) for $\alpha=\pi$, namely
when each bump plays the role of a cavity \cite{Phil90,schaff10},
corresponding to $v\gg c$ in our system.

\subsection{Random distribution of impurities}
\label{sec-num} 
As already illustrated in Sec. \ref{subsecthepot}, we consider
three sets of parameters for the AM and RDM potentials: (i) a bump
density $n_{\textmd{\textmd{dis}}}\simeq 0.12\lambda^{-1}\simeq 
0.16$ peaks/$\mu$m, and $d=\lambda$; (ii)
a bump
density $n_{\textmd{\textmd{dis}}}\simeq 0.5\lambda^{-1}
\simeq 0.65$ peaks/$\mu$m, and $d=\lambda$; and (iii)
with the same bump density as in (ii) but with $d=\lambda/2$.
In the three cases, the size of the disorder potential 
is $485$ $\mu$m and the BEC size is  204 $\mu$m.

We run the simulations for up to approximately 3-4 cycles of
$\omega_z$.  We generally note that in our range of drift velocities,
longer time durations are not necessary since the moving disorder
potential would try to bring the condensate too high up along the
harmonic potential and thus the condensate would inevitably fall back
at $z_f$.

\begin{figure}
\includegraphics[width=0.9\linewidth]{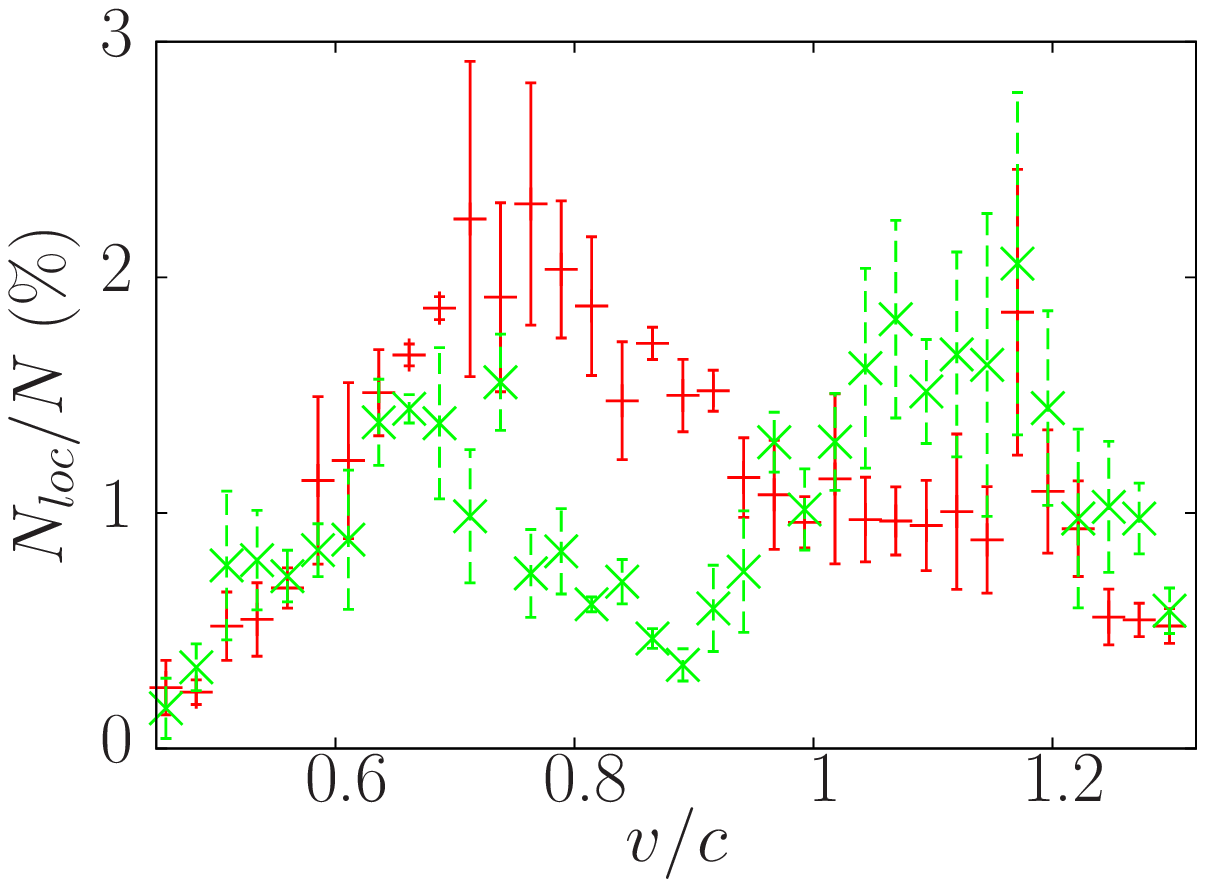}
\includegraphics[width=0.9\linewidth]{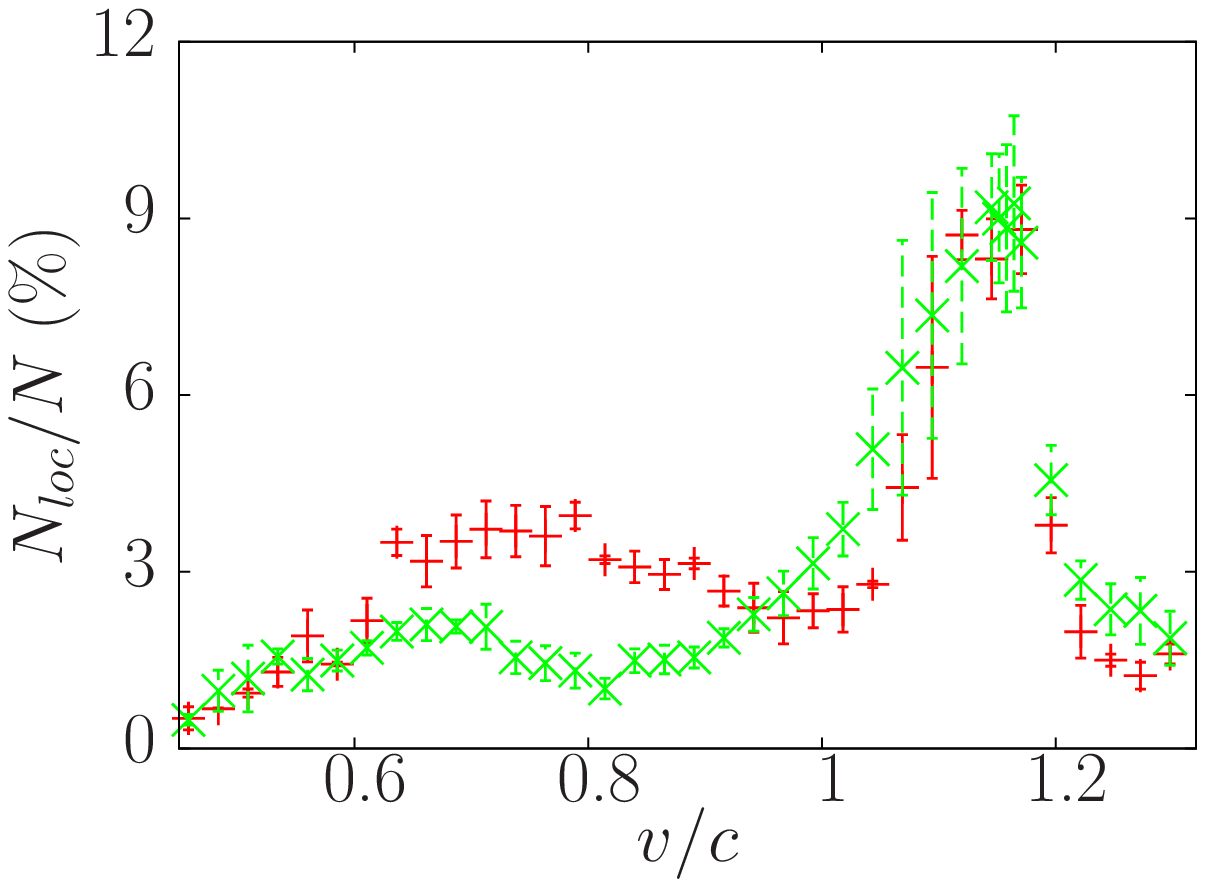}
\includegraphics[width=0.9\linewidth]{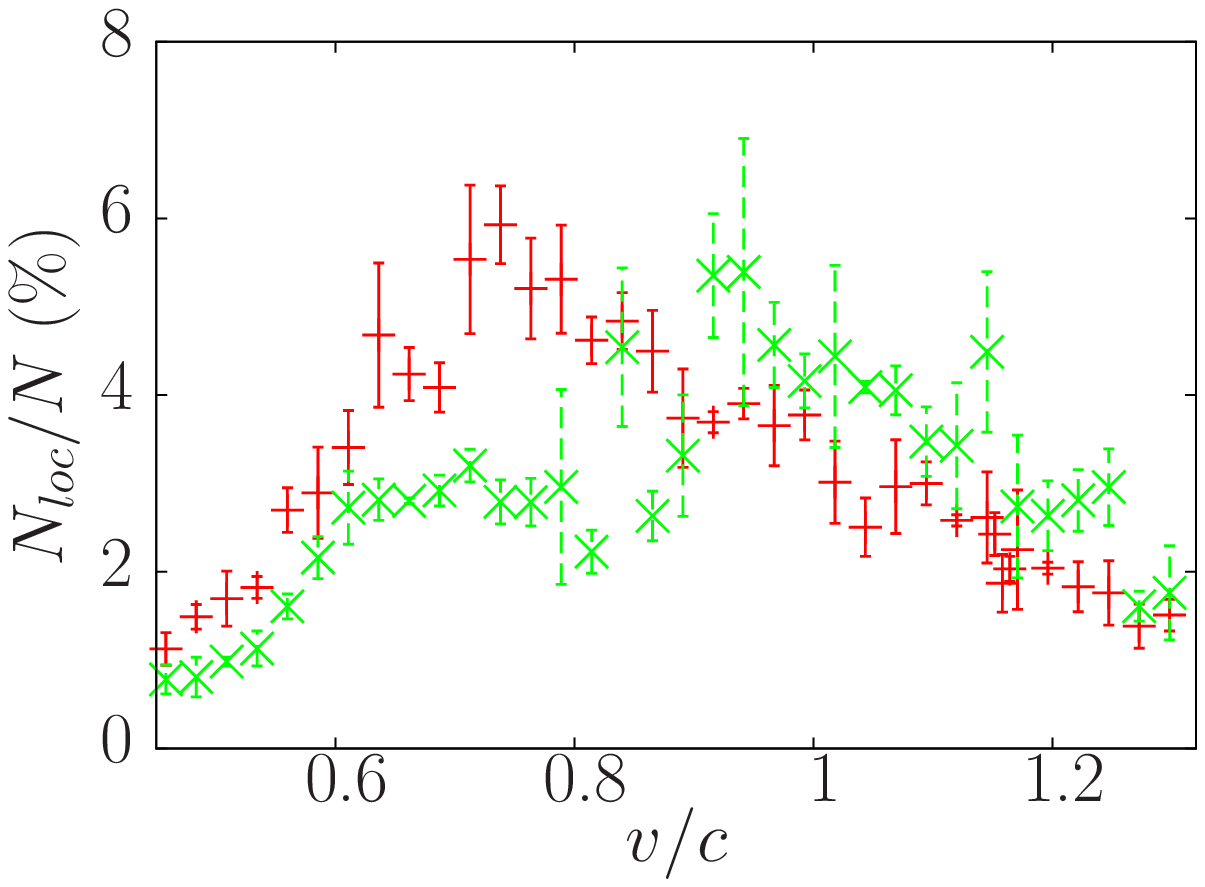}
\caption{(Color online) Localized BEC fraction as a function of the disorder potential 
drift velocity $v$, in units of the sound speed $c$ evaluated at the center of 
the trap. The red crosses correspond to the AM and the 
green stars to the RDM. The top panel corresponds to
$n_{\textmd{\textmd{dis}}}\simeq 0.12\lambda^{-1}$, and
$d=\lambda$; the middle panel corresponds to
$n_{\textmd{\textmd{dis}}}\simeq 0.5\lambda^{-1}$, and
$d=\lambda$; the bottom panel corresponds to
$n_{\textmd{\textmd{dis}}}\simeq 0.5\lambda^{-1}$, and
$d=\lambda/2$.
\label{figlocfrac}}
\end{figure}

We run simulations for 37 different drift velocities, in the range
$v=[0.46-1.30]c=[1.40-3.98] \textmd{mm/s}$ (where
$c=3.06\textmd{mm/s}$), with between 3 and 24 random potential
realizations for each drift velocity.  To obtain the percentage of
localized atoms, we compute the distance traveled by the condensate
center-of-mass corresponding to a fixed value of the potential
drift. We used that unique velocity-independent distance, in our case
$\Delta z_{\textmd{dis}}=96 \mu$m, in order to compare equivalent
final potential configurations.  For this value of $\Delta
z_{\textmd{dis}}$, we measure a small center-of-mass displacement
$\Delta z_{\mathrm{c.m.}}$, generally lower than an oscillator
harmonic length in the axial direction, $\Delta
z_{\mathrm{c.m.}}<a_z=\sqrt{\hbar/(m\omega_z)}$.  Then we identify the
ratio $N_{loc}/N$ with the ratio between the center-of-mass shift and
$\Delta z_{\textmd{dis}}$ according to Eq.\
(\ref{nloc}). Figure~\ref{figlocfrac} shows the localized ratio
$N_{loc}/N$ for the potentials (i), (ii) and (iii).  
We first note that irrespective of the kind of disorder the localization occurs
for $v>0.5c$. This unexpected localization
regime may be due to the local sound velocity being inhomogeneous due
to the harmonic confinement~\cite{Capuzzi2011} or to the non-vanishing
amplitude of the disorder potential~\cite{Ianeselli2006}.
$N_{loc}/N$ is very small for all data sets, except for the parameters
(ii) at $v\simeq 1.1 c$, for both AM and RDM (see the middle plot in
Fig.  \ref{figlocfrac}). We claim that this is an effect related to
the presence of dimers both in the RDM by construction, and in the AM
due to the choice $d=\lambda$ and density
$n_{\mathrm{dis}}=0.5\lambda^{-1}$, as shown in the bottom panel of
Fig. \ref{fig-autocorr}. Indeed, in the low density case (top panel of
Fig. \ref{figlocfrac}) where the AM has just developed few dimers with
respect to the RDM, the localization enhancement in the RDM is well
visible, even if the localization fraction is very small.  Moreover,
its oscillatory behavior as function of $v$ reminds that of the single
dimer shown in Fig.\ \ref{fig-singledef}, with minimal localization
near $v/c\simeq 0.9$.  Finally, in the bottom panel of
Fig. \ref{figlocfrac} we show that the choice $d=\lambda/2$ suppresses
the $N_{loc}/N$ peak at $v\simeq 1.1 c$, since it destroys the
correlations introduced by the single dimer itself
(cf. Fig. \ref{fig-autocorr2}).
 
\begin{figure}
\includegraphics[width=0.9\linewidth]{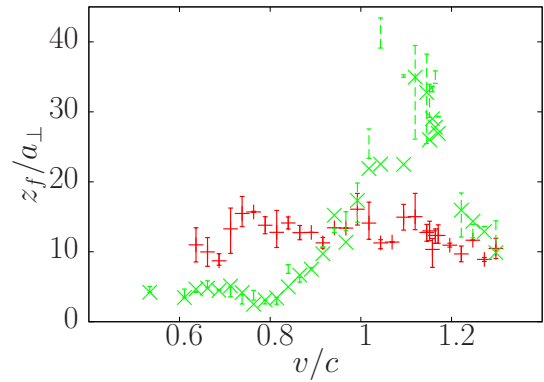}
\caption{(Color online) Turning point $z_f$ as a function of the
disorder potential velocity $v$, in units of the sound speed $c$
evaluated at the center of the trap.  The red crosses correspond
to the uncorrelated disorder with
$n_{\mathrm{dis}}=0.5\lambda^{-1}$ and $d=\lambda/2$, and the
green stars to the correlated one with $d=\lambda$ and the same
peak density.  \label{figturningpoint}}
\end{figure}

In Fig.~\ref{figturningpoint} we plot the turning point $z_f$ for the
AM with $n_{\textmd{dis}}=0.5\lambda^{-1}$ and $d=\lambda/2$ and the
RDM with the same peak density but with $d=\lambda$.  We verified
numerically that for the disorder amplitude and velocities considered
the turning point $z_f$ is proportional to the average drag force
$\bar F_{\textmd{dis}}$ as expected from Eq. (\ref{eq:turningpoint})
in the limit of small displacements.  Not surprisingly, the overall
behavior of the average drag force $\bar F_{\textmd{dis}}$ is
qualitatively similar to the localization ratio, showing that to a
greater drag force, i.e., less superfluid fraction, it corresponds a higher
localization efficiency.  In agreement with the single defect analysis
(Sec.  \ref{sec-corr}), in both Figs.~\ref{figlocfrac} and
\ref{figturningpoint}, we clearly observe a partial suppression of the
localization at $v\simeq 0.8-0.9 c$, and an enhancement of the
localization at a supersonic region of the drift velocities, when
correlations are dominant.


\subsection{Localization inhomogeneity}
\begin{figure}
\includegraphics[width=0.9\linewidth]{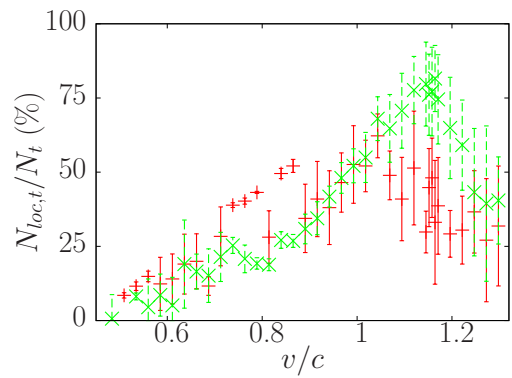}
\caption{(Color online) Localized BEC fraction of the leading moving
  tail -1\% of the total condensate mass- as a function of the disorder
potential velocity $v$, in units of the sound speed $c$ evaluated at the center of the
trap. The red crosses correspond to the AM (iii) and the green stars to the RDM (ii).
\label{locfractaildimer1per}}
\end{figure}

In order to prove that we are observing the localization of a BEC
fraction rather than a slower drag of the whole BEC, we analyze the
dynamics of the BEC tails, where the density is lower and AL should
occur more efficiently.  In particular, we focus on the forward moving
tail distribution that experiences the disorder potential for the
whole simulation time.  Analogously to Sec. \ref{sec-num}, we compute
the percentage of atoms localized in the forward moving tail of the
condensate $N_{loc,t}/N_t$ by dividing the tail center-of-mass shift
with the same $\Delta z_{\textmd{dis}}$ value. We looked at 3
different sizes of that tail: those comprising 1\%, 5\% and 10\% of
the total condensate mass.  In the case of the forward tails that
amounts to 1\% of the total condensate mass (see
Fig. \ref{locfractaildimer1per}), we observe an almost complete
localization in that same supersonic region of the drift
velocities. While for the 5\% and 10\% mass tails (not shown), we
measure $~$ 50\% of localization efficiency.  The negative values of
localization in the subsonic region of drift velocities are due to the
fluctuations of the condensate density caused by the presence of the
disorder potential. At those velocities, these fluctuations provoke
the forward leading small portion of mass to buckle back towards the
center of the condensate even though there is no much overall motion
of the whole condensate.

\section{Conclusions}
\label{concl}
We have presented an analysis of the drag properties of two kinds of
disorder potentials of finite extent moving through an inhomogeneous
quasi-1D Bose-Einstein condensate. Because of the presence of the
external harmonic confinement our system, unlike \cite{leboeuf09}, is
never in a stationary state. We treated both cases of non-correlated
and correlated disorder with short-range correlations.  Our numerical
computation of the fraction of localized atoms and the drag force shows
that for the case of correlated disorder, in the form of random
dimers, there is a suppression or an enhancement of localization 
depending on the drift velocity. This was buttressed by our analytical
optical model in which we determined the reflectivity of a single
dimer and of a lonely defect.  
The effects of correlations are masked as we increase the disorder
density and the dimers cannot be distinguished from a high density
collection of single defects.

\begin{acknowledgments}
This work was supported by CNRS PICS grant No. 05922.  A. A. is
grateful to M. Albert for fruitful discussions.  P. C.
acknowledges support  from ANPCyT and CONICET, Argentina through
grants PICT 2008-0682 and PIP 0546, respectively.
\end{acknowledgments}

\end{document}